\documentclass[sigconf]{acmart}

\usepackage[ruled,vlined,linesnumbered]{algorithm2e}
\SetKwInput{KwParam}{Parameter}
\usepackage{subcaption}
\usepackage{multirow}
\usepackage{pifont}
\usepackage{enumitem}
\usepackage{bm}

\AtBeginDocument{
  }

\copyrightyear{2025}
\acmYear{2025}
\setcopyright{cc}
\setcctype{by}
\acmConference[MM '25] {Proceedings of the 33rd ACM International Conference on Multimedia}{October 27--31, 2025}{Dublin, Ireland.}
\acmBooktitle{Proceedings of the 33rd ACM International Conference on Multimedia (MM '25), October 27--31, 2025, Dublin, Ireland}
\acmISBN{979-8-4007-2035-2/2025/10}
\acmDOI{10.1145/3746027.3761986}

\graphicspath{{./images/}} 

\begin{document}

\title{MIG-COW: Transferable Adversarial Attacks on Deepfake Detectors via Gradient Decomposition}

\author{Wonjune Seo}
\authornote{Equal contribution}
\orcid{0009-0008-8841-2849}
\affiliation{%
  \institution{Ulsan National Institute of Science and Technology}
  \city{Ulsan}
  \country{Republic of Korea}}
\email{wonjuneseo@unist.ac.kr}

\author{Joonhyuk Baek}
\authornotemark[1]
\orcid{0009-0006-7470-0255}
\affiliation{%
  \institution{Ulsan National Institute of Science and Technology}
  \city{Ulsan}
  \country{Republic of Korea}}
\email{jjhk7330@unist.ac.kr}

\author{Yeseong Jung}
\authornotemark[1]
\orcid{0009-0004-3240-3442}
\affiliation{%
  \institution{Ulsan National Institute of Science and Technology}
  \city{Ulsan}
  \country{Republic of Korea}}
\email{y2ahsong@unist.ac.kr}

\author{Saerom Park}
\authornote{corresponding author}
\orcid{0000-0002-2687-7105}
\affiliation{%
  \institution{Ulsan National Institute of Science and Technology}
  \city{Ulsan}
  \country{Republic of Korea}}
\email{srompark@unist.ac.kr}

\renewcommand{\shortauthors}{Wonjune Seo, Joonhyuk Baek, Yeseong Jung, and Saerom Park}

\begin{abstract}
Despite recent advances, deepfake detectors remain vulnerable to adversarial examples, particularly in diverse, real-world settings. We propose MIG-COW, a novel adversarial attack framework that generates highly generalizable and visually imperceptible adversarial examples. By combining momentum-integrated gradients with a consensus-orthogonal decomposition, MIG-COW captures both shared and model-specific vulnerabilities across heterogeneous CNN and ViT detectors. On the AADD-2025 Challenge benchmarks, MIG-COW achieves a 99.96\% white-box attack success rate (ASR) with high perceptual similarity (SSIM), significantly outperforming existing baselines. However, its limited 7.16\% ASR against official black-box targets—despite achieving the best overall score—highlights the ongoing challenge of transferability. We also demonstrate that incorporating low-performing but diverse models in the ensemble can degrade attack effectiveness, underscoring the need for careful surrogate model selection in real-world adversarial settings.
\end{abstract}



\maketitle
\section{Introduction}
Deepfake detection models have been widely integrated into modern security frameworks, often demonstrating impressive accuracy and robustness under conventional operating conditions. Nevertheless, a critical question remains: how resilient are these systems in the presence of sophisticated adversarial threats? If a deepfake image---once confidently identified as ``face''---can be subtly manipulated to evade detection and be misclassified as ``real,'' it reveals a fundamental vulnerability in current detection systems. This highlights a significant challenge that current detection technologies have not yet fully addressed.

Most state-of-the-art deepfake detectors can be broadly categorized into two architectural paradigms: convolutional neural network (CNN)-based models, which  capture local texture and artifact patterns, and vision transformer (ViT)-based models, which use self-attention to model global context. Despite their success, both  remain susceptible to adversarial attacks. In particular, CNNs possess strong local structural inductive biases, whereas ViTs are intentionally designed to reduce such biases, relying more on data-driven learning of relationships. Consequently, adversarial perturbations often transfer poorly between these two distinct architectures~\cite{bhojanapalli2021understanding, guesmi2025tesser}.

This observation motivates a more fundamental question: \textbf{Can we generate a single adversarial example capable of simultaneously deceiving multiple detection architectures?} Addressing this question requires uncovering the underlying vulnerabilities that are common across diverse models while minimizing reliance on architecture-specific components such as direct gradient with respect to input. To address it, we propose a novel ensemble attack framework, \textbf{MIG-COW} (Momentum-Integrated Gradient with Consensus-Orthogonal Weighting), designed to enhance the generalizability of adversarial examples in deepfake detection without compromising white-box attack performance. By systematically identifying and disentangling both shared and distinct decision patterns, we can more effectively evaluate the true resilience of both existing and unknown detectors under realistic adversarial conditions.

In this work, we make the following key contributions:
\begin{itemize}[topsep=0pt, partopsep=0pt]
    \item We propose MIG-COW, a novel ensemble adversarial attack framework that generates highly generalizable and visually imperceptible adversarial examples by effectively exploiting both shared and model-specific vulnerabilities across CNN and ViT-based deepfake detectors.
     \item We conduct extensive experiments across diverse white-box and black-box detectors, comparing against existing adversarial attack methods as baselines. Our proposed method achieves the highest overall score—balancing both attack success and visual imperceptibility—across two white-box and two black-box models. 
    \item We provide novel insights into transfer-based attacks, revealing that incorporating low-performing source models can degrade overall attack effectiveness despite architectural diversity, underscoring the critical importance of functional equivalence for robust adversarial transfer.
\end{itemize}

\section{Related work}
\label{sec:related_work}
Recent research on deepfake detection can be broadly categorized into CNN-based and ViT-based approaches.
CNN-based detectors focus on local textures and manipulation artifacts, with models like Xception~\cite{Rossler_2019_ICCV} and EfficientNet-B7~\cite{seferbekov2020dfdc} performing well on datasets such as FaceForensics++ and DFDC.
In contrast, ViT-based detectors capture global dependencies using self-attention.
The ICT\cite{Dong2022CVPR} enforces identity consistency across video frames, while UIA-ViT\cite{zhuang2022uia} detects intra-frame forgery signals in an unsupervised manner.

Despite the progress made by both architectural paradigms, significant challenges remain. Their inherent structural differences that exhibit different inductive biases lead to varying vulnerabilities to adversarial attacks and limited generalizability to unseen manipulations or domain shifts.

Gradient-based adversarial attacks are widely used to test model robustness.
FGSM~\cite{goodfellow2014explaining} and PGD~\cite{madry2017towards} serve as baseline white-box attacks. To enhance transferability, MI-FGSM~\cite{dong2018boosting} leverages momemtum-based updates that stabilizes gradient updates and achieves higher success rates against unseen models.

Another direction for enhancing transferability involves using attribution-based gradients, which offer more semantically meaningful perturbation directions. 
MIG~\cite{ma2023transferable} introduced integrated gradients with momentum to produce more architecture-invariant perturbations, effectively targeting both CNNs and ViTs. 
More recently, MuMoDIG~\cite{ren2025improving} refined the integration path and demonstrated that using a log-based loss instead of a raw output loss further enhances the effectiveness of the attack.

Nevertheless, few studies have explicitly investigated the performance of these general-purpose gradient-based attacks on deepfake detection models. Unlike standard classifiers, deepfake detectors often depend on subtle cues such as fine-grained textures and identity consistency, making them qualitatively different. 
This gap highlights the need for tailored attack strategies that account for the heterogeneous nature of deepfake detection systems and can expose shared vulnerabilities without sacrificing attack effectiveness on individual models across diverse architectures.

\section{Methodology}
\subsection{Overview}
\begin{figure*}[t]
    \centering
    \includegraphics[width=0.95\textwidth]{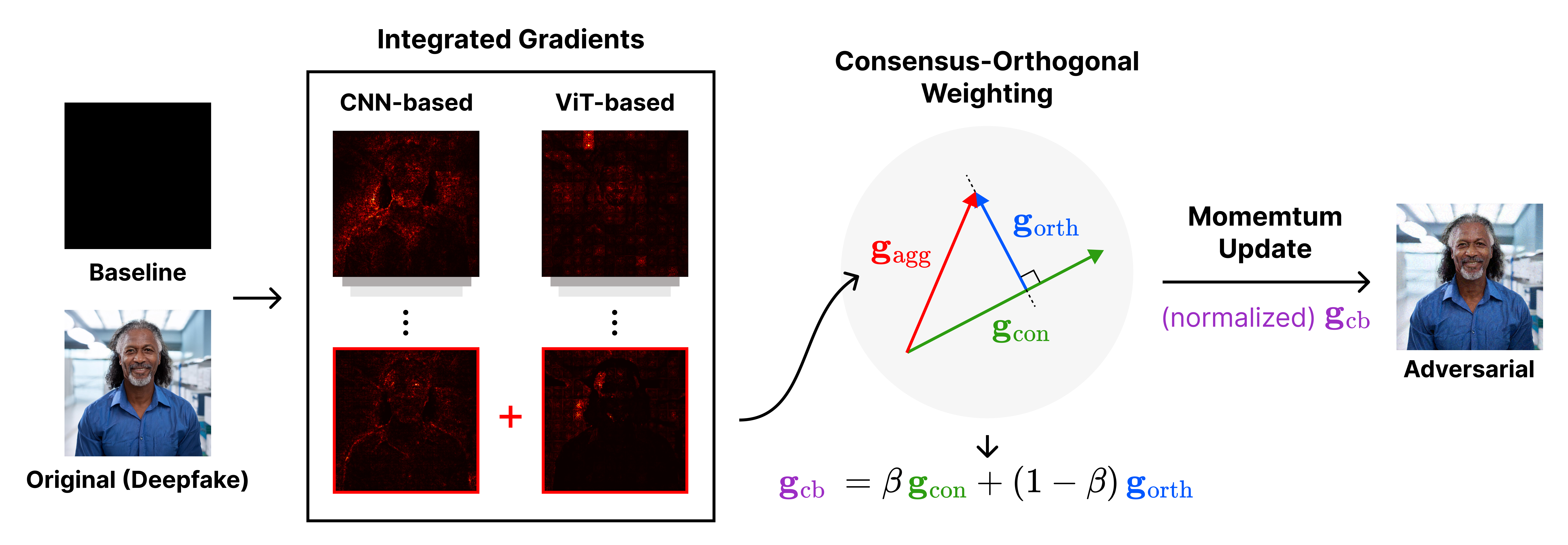}
    \caption{Overview of the proposed MIG-COW framework. The attack integrates momentum-based integrated gradients and consensus-orthogonal weighting to generate transferable adversarial examples across diverse deepfake detectors.}
    \label{fig:overview}
\end{figure*}

We propose an attack framework called \textbf{Momentum Integrated Gradient with Consensus-Orthogonal Weighting (MIG-COW)} to generate adversarial examples with enhanced generalizability across a wide range of deepfake detection models, including both CNNs and ViT. An overview of the proposed MIG-COW framework is illustrated in Figure~\ref{fig:overview}.

Our method leverages two key properties to promote the generalization of adversarial perturbations from \textit{source models} (known detection models) to \textit{target models} (unseen detection models).
\begin{itemize}
    \item Implementation invariance~\cite{sundararajan2017axiomatic, ma2023transferable,ren2025improving}: We adopt Integrated Gradients (IG), which are invariant to the model’s implementation details and depend only on the input-output mapping. Since adversarial transferability is more likely when source and target models share similar functional behavior, leveraging IG improves attack generalization across architectures.
    \item Gradient Ensemble~\cite{ma2023transferable}: As highlighted by \cite{ma2023transferable}, ensembling gradients from multiple models or inputs is known to enhance the robustness and generalizability of adversarial attacks. Specifically, combining IG or logit gradients yields better results than directly aggregating raw perturbations. 
\end{itemize}

A critical challenge in gradient ensembling—especially in achieving cross-model generalizability—is how to aggregate gradients in a way that preserves attack effectiveness on source models. To address this, we propose a novel aggregation strategy that combines consensus gradient directions with complementary architecture-specific information. The resulting MIG-COW framework consists of two main components: (1) momentum-integrated gradients and (2) gradient aggregation via orthogonal decomposition.

\subsection{Proposed Method}
\subsubsection{Integrated Gradient (IG)}
\begin{figure}[htbp]
    \centering
    \includegraphics[width=0.90\linewidth]{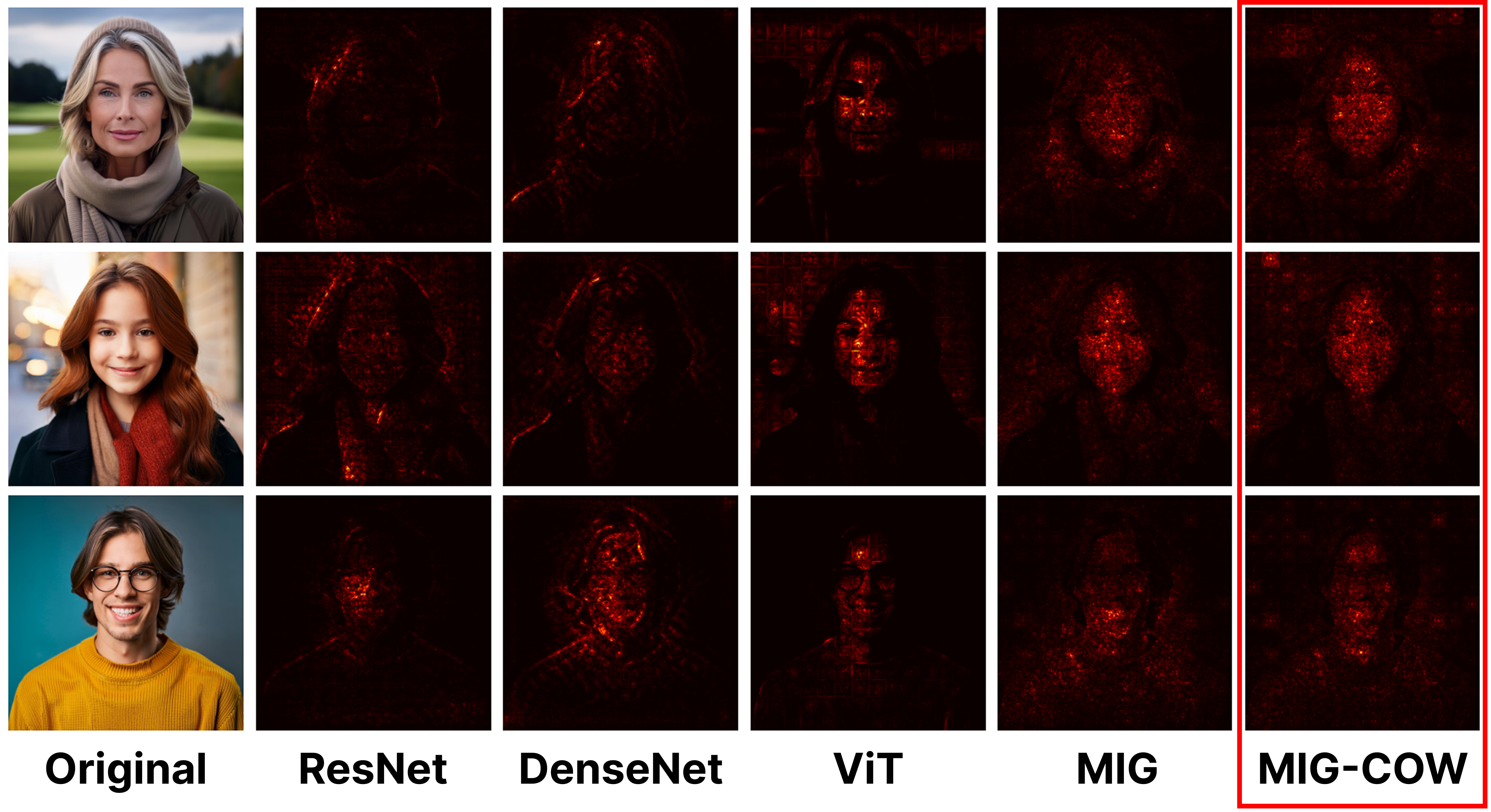}
    \caption{Comparison of IG visualizations across different model: ResNet50, DenseNet121, ViT-P, MIG, and our proposed method.}
    \label{fig:igmap}
\end{figure}

In our proposed method, we compute IG for each source model with respect to the target label (i.e., ``real'' class in deepfake detection) to identify which input features contribute most to the prediction. IG computes the path-integrated gradient from a black baseline image to the input, producing stable and interpretable attributions across interpolation steps. This approach is beneficial for both CNNs and ViTs due to its robustness to model-specific variations. Figure~\ref{fig:igmap} shows the IG attribution maps of individual models, MIG, and our proposed method (MIG-COW). By comparing the attribute maps, we can see how the proposed method improves transferability. In particular, our method effectively captures the core areas common to different models and the unique contributions of each architecture.

For a given input $x\in \mathcal{X}$, baseline $b\in \mathcal{X}$ (i.e., a black image), and source model $f: \mathcal{X}\rightarrow [0,1]$, the IGs for the $i$-th input dimension are defined as:
\begin{equation}
    \mathbf{g}_i = IG_i(f, x, b) = (x_i - b_i) \times \int_{\xi=0}^1 
    \frac{\partial f\big( b + \xi (x - b) \big)}{\partial x_i} d\xi.
    \label{eq:ig}
\end{equation}
In practice, this path integral is approximated by a Riemann sum with $s$ discrete steps:
\begin{equation}
    IG_i(f, x, b) \approx (x_i - b_i) \times 
    \sum_{k=1}^s 
    \frac{\partial f\big( b + \frac{k}{s} (x - b) \big)}{\partial x_i} \times \frac{1}{s}.
    \label{eq:ig-approx}
\end{equation}

We also employ momentum-integrated gradients (MIG)~\cite{ma2023transferable} to stabilize the optimization process and help escape undesirable local optima by updating the accumulated gradient with a momentum term at each iteration. 
To enhance generalizability and gradient scaling, inspired by ~\citet{ren2025improving}, we define the source model $f(x)$ in Eq. ~\ref{eq:ig} and Eq. ~\ref{eq:ig-approx} as the cross-entropy loss computed directly on the raw logits for the target class, rather than using the raw probability. This implicitly leverages the log-probability, enhancing gradient scaling and resulting in more robust gradients without requiring extra softmax normalization.

Furthermore, we decompose the gradient space into two complementary components: a consensus direction that averages L2-normalized attribution maps to capture shared vulnerabilities, and an orthogonal direction derived from the least-aligned eigenvector of the gradient Gram matrix to remove redundant correlations. The final attack direction is then constructed as a weighted combination of these two components. 

\subsubsection{Consensus Orthogonal Weighting}
Given $N$ source models, let $\{\mathbf{g}_i\}_{i=1}^N$ denote the IGs obtained from each model, where each $\mathbf{g}_i \in \mathbb{R}^{C \times H \times W}$ has the same shape as the input image $x \in \mathcal{X} = \mathbb{R}^{C \times H \times W}$. The consensus gradient $\mathbf{g}_{\text{con}}$ is then obtained by ensembling the individual IGs:
\begin{equation}
\mathbf{g}_{\text{con}} = \frac{1}{N} \sum_{i=1}^N \mathbf{g}_i.
\end{equation}
However, relying solely on this direct averaging approach may overemphasize redundant or shared information while obscuring unique discriminatory information that might be present in individual model gradients. 

\noindent\textbf{Compute model-specific directions.} To analyze the relationships among the gradients, we construct a Gram matrix $\bm{K}$. We first flatten each gradient $\mathbf{g}_i$ into a vector $\tilde{\mathbf{g}}_i = \text{vec}(\mathbf{g}_i)\in \mathbb{R}^D$ (where $D=C\times H\times W$), then stack these vectors to obtain a matrix $\bm{G}$, and compute the Gram matrix $\bm{K}$:
\begin{equation}\label{eq:gram}
    \bm{G} = [\tilde{\mathbf{g}}^{(1)}, \ldots, \tilde{\mathbf{g}}^{(N)}]\in \mathbb{R}^{D\times N}, \quad \bm{K}=\bm{G}^\top \bm{G} \in \mathbb{R}^{N\times N},
\end{equation}  
where each entry $\bm{K_{ij}}$ corresponds to the inner product $\langle \tilde{\mathbf{g}}_i, \tilde{\mathbf{g}}^{(j)} \rangle$, capturing the degree of correlation between the gradients from models $i$ and $j$. This matrix $\bm{K}$ helps to identify the correlation structure among the ensemble gradients.

To extract the least aligned direction relative to the consensus $\mathbf{g}_{\text{con}}$, we perform eigen decomposition of the Gram matrix and select the eigenvector $\bm{v_{\min}}$ corresponding to the smallest eigenvalue $\lambda_{\min}$, such that $\bm{K} \bm{v_{\min}} = \lambda_{\min} \bm{v_{\min}}$.
This eigenvector captures the direction with minimum redundancy across model gradients—highlighting model-specific or orthogonal components that are underrepresented in the consensus. Using $\bm{v_{\min}}$, we construct the aggregated gradient $\mathbf{g}_{\text{agg}}$ through its smallest eigenvector as a weighted sum of IGs:

\begin{equation}
    \mathbf{g}_{\text{agg}} = \sum_{i=1}^N v_{\min}^{(i)}\, \mathbf{g}_i,
\end{equation}
where $v_{\min}^{(i)}$ is the $i$-th entry of the eigenvector $\bm{v_{min}}$. 

However, since the consensus vector $\mathbf{g}_\text{con}$ and the aggregated vector $\mathbf{g}_{\text{agg}}$ may still share redundant information, we remove the component of $\mathbf{g}_{\text{agg}}$ that lies in the direction of $\mathbf{g}_{\text{con}}$ to retain only the orthogonal and non-redundant parts as follows:
\begin{equation}
    \mathbf{g}_{\text{orth}} = \mathbf{g}_{\text{agg}} - \frac{\langle \mathbf{g}_{\text{agg}}, \mathbf{g}_{\text{con}} \rangle}{\|\mathbf{g}_{\text{con}}\|^2 + \delta} \, \cdot\mathbf{g}_{\text{con}}.
\end{equation}
As a result, $\mathbf{g}_{\text{orth}}$ captures the model-specific directions and enhances adversarial transferability.

\noindent\textbf{Final Attack Direction.} 
Finally, we construct the combined direction for the attack by weighting the consensus vector and the orthogonal vector as follows:
\begin{equation}\label{eq:comb}
    \mathbf{g}_{\text{cb}} = \beta \, \mathbf{g}_{\text{con}} + (1-\beta) \, \mathbf{g}_{\text{orth}},
\end{equation}
where $0<\beta<1$. We also normalize the gradient~\eqref{eq:comb}.

\subsubsection{Perform adversarial update with momentum.} At each iteration, the adversarial example is updated with a momentum term:
\begin{equation}
    \mathbf{g}_{(t)} = \mu \, \mathbf{g}_{(t-1)} + \frac{\mathbf{g}_{\text{cb}}}{\|\mathbf{g}_{\text{cb}}\|}, 
    \quad 
    \boldsymbol{x}_{(t)} = \text{Clip}_\epsilon \big( \boldsymbol{x}_{(t-1)} + \alpha \cdot \text{sign}(\mathbf{g}_{(t)}) \big),
\end{equation}
where $\mu$ is the momentum decay factor and $\epsilon$ is the perturbation budget.

This design allows our attack to benefit from both the robust consensus direction and model-specific diverse cues, thus boosting generalizability across different architectures.
Algorithm~\ref{alg:algo_sim} details the overall procedure of MIG-COW.

\begin{algorithm}
\caption{MIG-COW: Momentum Integrated Gradients with Consensus-Orthogonal Weighting}
\label{alg:algo_sim}
\KwIn{
  White-box source models $\{ f_i \}_{i=1}^N$, original clean image $\boldsymbol{x}$ and target label $y$.
}

\KwParam{
  Perturbation budget $\epsilon$, iteration number $T$, momentum factor $\mu$, weight factor $\beta$, baseline image $\boldsymbol{b}$. Small constant $\delta$ for numerical stability\;
}

\KwOut{
  Adversarial image $\bm{x_{\text{adv}}}$.
}

Initialize accumulated gradient $\mathbf{g}_{(0)} \leftarrow \mathbf{0}$\;
Set step size $\alpha \leftarrow \epsilon / T$, and $\boldsymbol{x}_{(0)} \leftarrow \boldsymbol{x}$\;

\For{$t = 1$ to $T$}{

  \tcp{Integrated Gradients for each model}
  \For{$i = 1$ to $N$}{
  $\mathbf{g}_i \leftarrow IG(f_i, \boldsymbol{x}_{(t-1)}, \boldsymbol{b})$\;
  }
    $\mathbf{g}_{\text{con}} \leftarrow \frac{1}{N} \sum_{i=1}^N \mathbf{g}_i$\;

  \tcp{Flattened gradients and Gram matrix}
  $\mathbf{G} \leftarrow [\text{vec}(\mathbf{g}_1), \ldots, \text{vec}(\mathbf{g}_N)] \in \mathbb{R}^{D \times N}$\;
  
  $\mathbf{K} \leftarrow \bm{G^T} \bm{G}$\;

  \tcp{Compute the least-redundant direction}
  Compute eigenvector $\bm{v_{\min}}$ of $\bm{K}$ with smallest eigenvalue\;
  $\mathbf{g}_{\text{agg}} \leftarrow \sum_{i=1}^N v_{\min}^{(i)} \, \mathbf{g}_i$
   
  $\mathbf{g}_{\text{orth}} \leftarrow \mathbf{g}_{\text{agg}} - \frac{\langle \mathbf{g}_{\text{agg}}, \mathbf{g}_{\text{con}} \rangle}{\| \mathbf{g}_{\text{con}} \|^2 + \delta} \cdot\mathbf{g}_{\text{con}}$\;
  
  \tcp{Final attack direction}
  
    $\mathbf{g}_{\text{cb}} \leftarrow \beta \cdot \mathbf{g}_{\text{con}} + (1 - \beta) \cdot \mathbf{g}_{\text{orth}}$\\
    $\mathbf{g}_{(t)} \leftarrow \mu \cdot \mathbf{g}_{(t-1)} + \frac{\mathbf{g}_{\text{cb}}}{\| \mathbf{g}_{\text{cb}} \|_1}$\;
    
  \tcp{Update adversarial image}
  
    $\bm{x}_{(t)} = \text{Clip}_\epsilon \big( \bm{x_{(t-1)}} + \alpha \cdot \text{sign}(\mathbf{g}_{(t)}) \big)$\;
}

$\boldsymbol{x}_{\text{adv}} \leftarrow \boldsymbol{x}_{(T)}$\;

\KwRet{$\bm{x_{\text{adv}}}$}
\end{algorithm}

\begin{table*}[t]
  \centering
  \renewcommand{\arraystretch}{1.2}
  \begin{tabular}{l|l|ccc|c|cc|c|c|c}
    \toprule
    \multirow{2}{*}{Setting} 
    & \multirow{2}{*}{Method} 
    & \multicolumn{4}{c|}{White-box ASR (\%)} 
    & \multicolumn{3}{c|}{Black-box ASR (\%)} 
    & \multirow{2}{*}{SSIM ↑} 
    & \multirow{2}{*}{Score} \\
    \cmidrule(lr){3-6} \cmidrule(lr){7-9}
    & & ResNet50 & DenseNet121 & ViT-P & Avg & ViT-B/16 & DenseNet121-DCT & Avg & & \\
    
    \midrule
    \multirow{4}{*}{w/o ViT-P} 
    & PGD & 56.45 & 60.37 & - & 58.31 & 0.50 & 7.98 & 4.02 & \textbf{0.9304} & 1655 \\
    & MI-FGSM & 88.03 & 93.94 & - & 91.00 & 3.42 & 9.10 & 6.26 & 0.9096 & 2485 \\
    & MIG & 96.94 & 90.02 & - & \underline{93.48} & 3.64 & 9.90 & \underline{6.77} & \underline{0.9169} & \underline{2577} \\
    & MIG-COW & 100 & 99.93 & - & \textbf{99.96} & 3.50 & 10.83 & \textbf{7.16} & 0.9143 & \textbf{2747} \\
    
    \midrule
    \multirow{4}{*}{w/ ViT-P} 
    & PGD      & 57.02 & 63.64 & 25.74 & 48.80 & 0.64  & 9.10  & 4.87    & 
    \textbf{0.9280} & 1716 \\
    & MI-FGSM &  86.74 & 92.59 & 51.53 & 76.95 & 3.06 & 12.12 & \textbf{7.59} & 0.9166 & 2480 \\
    & MIG      & 94.80 & 85.79 & 90.96 & \underline{90.52} & 2.00 & 8.70  & 5.35    & \underline{0.9182}   & \underline{2464} \\
    & MIG-COW  & 99.93 & 99.93 & 93.87 & \textbf{97.91} & 1.50 & 9.70 & \underline{5.60} & 0.9161 & \textbf{2709} \\
    \bottomrule
    
  \end{tabular}
  \caption{Comparison of attack methods on white-box and black-box detectors, including average ASR, SSIM, and overall score, where the best results are highlighted in bold, and the second-best results are underlined. The Score is computed based on the four models excluding ViT-P (i.e., ResNet50, DenseNet121, ViT-B/16, and DenseNet121-DCT).}
  \label{tab:full_comparison}
\end{table*}

\section{Experiments}
\subsection{Datasets and Evaluation}
We evaluate the effectiveness of MIG-COW using the dataset provided by the AADD-2025 organizers. The dataset is divided into HQ and LQ subsets. Specifically, the HQ subset consists of 693 fake images generated by various models, including Adobe Firefly~\cite{adobe_firefly}, DeepAI~\cite{deepai_text2img}, Flux 1.1 Pro~\cite{bfl_flux_repo}, HotPotAI~\cite{hotpotai}, Nvidia SanaPAG~\cite{xie2024sana}, Stable Diffusion 3.5 \cite{stabilityAI_sd35_code}, StyleGAN 2~\cite{karras2020analyzing}, StyleGAN 3~\cite{karras2021alias}, and Tencent Hunyuan~\cite{li2024hunyuan}. The LQ subset includes 710 fake images created using DeepAI, Flux 1.1~\cite{bfl_flux_repo}, Freepik~\cite{freepik}, HotPotAI, Nvidia SanaPAG Stable Diffusion Attend and Excite~\cite{chefer2023attend}, StyleGAN \cite{karras2019style}, StyleGAN 3, and Tencent Hunyuan. LQ images are produced by resizing the originals and applying variable Quality Factor compression to simulate the degradation typically observed in social media environments.

All 1,403 fake images were verified in advance to be correctly classified as fake by the two white-box detection models provided by the organizers. We then generate adversarial examples for these images using MIG-COW and evaluate the attack success rate (ASR). Specifically, we calculate the white-box ASR by counting the number of images that successfully deceive the white-box models and dividing this by the total number of images. Similarly, we compute the black-box ASR by evaluating how many adversarial examples successfully fool unseen black-box detectors. Additionally, to assess the visual quality of the adversarial examples, we measure their similarity to the original images using SSIM.

Furthermore, we follow the official evaluation criteria provided by the AADD-2025 challenge. The score is computed based on a combination of visual similarity and attack effectiveness. 
The score is calculated as:
\begin{equation}
\sum_{C_f \in \mathcal{C}} \sum_{k=1}^{n_{\text{test}}} SSIM(I_k, I_k^{ADV}) \cdot \mathbb{I}\left[ C_f(I_k^{ADV}) = \text{LABEL}_{real} \right]
\end{equation}

where $\mathcal{C}$ denotes the set of evaluation classifiers, and $n_{\text{test}}$ is the number of test images. $I_k$ and $I_k^{ADV}$ represent the original and adversarial images, respectively. $\text{LABEL}_{real}$ refers to the ground-truth label for the real class. The indicator function $\mathbb{I}[\cdot]$ returns 1 if the classifier's prediction matches the real label, and 0 otherwise.

\subsection{Experiment Settings}
\label{sec:experiment_settings}
We evaluate the effectiveness of adversarial examples generated by various methods by measuring both their white-box and black-box ASRs. As discussed in Section~\ref{sec:related_work}, recent research in deepfake detection can be broadly categorized into CNN-based and ViT-based approaches. To reflect this architectural diversity and analyze differences in attack robustness, we adopt two CNN-based detectors—ResNet50~\cite{he2016deep} and DenseNet121~\cite{huang2017densely}—provided by the competition organizers as white-box source models, and additionally include a publicly available pretrained ViT-based model~\cite{dosovitskiy2020image, Deep-Fake-Detector-v2-Model} (referred to as ViT-P) as a supplementary white-box source. Since the ViT-P was not fine-tuned on the AADD-2025 deepfake dataset, it correctly classified only approximately 70\% of the 1,403 clean fake images as fake, reflecting a domain mismatch that degrades its baseline performance, unlike other white-box models.

For black-box evaluation, we use two official challenge models based on the ViT-B/16~\footnote{Although our supplementary white-box model (ViT-P) shares the same ViT-B/16 architecture, it differs significantly in weights and performance. Unlike ViT-P, the official black-box model achieves near-perfect accuracy on the AADD-2025 deepfake dataset.} and DenseNet121-DCT architectures.
Additional details on the black-box setting are provided in Section~\ref{sec:black_att}.

To provide a benchmark for the proposed MIG-COW, we compare its performance against widely used gradient-based attack methods, including PGD, MI-FGSM, and the base MIG framework. For these baseline methods, we adopt a simple ensemble strategy (averaging) as in~\cite{ren2025improving} to enhance their transferability and ensure a fair comparison with MIG-COW's multi-model design. For consistency across all experiments, we set the perturbation budget $\epsilon=0.02$, the number of iterations $T=25$, and the momentum factor $\mu=1$.

\subsection{Experimental Results}
\label{sec:exp_res}
Table~\ref{tab:full_comparison} summarizes results for both white-box and black-box attacks across two evaluation settings—based on whether the ViT-P is included among the white-box detectors.\footnote{We submitted results using ViT-P for the AADD-2025 challenge; however, after the official evaluation code was released, we find that excluding ViT-P led to improved overall performance.} For comparison, overall scores reported in Table~\ref{tab:full_comparison} are computed using only the two official white-box models and two black-box models provided by the AADD-2025 challenge. 
\subsubsection{Attack on White-box Detectors}
In the white-box setting, we evaluate the effectiveness of adversarial examples when the attacker has full knowledge of the deepfake detector's architecture and parameters.  

In the first setting, which considers only CNN-based detectors (ResNet50 and DenseNet121), gradient-based attacks such as PGD and MI-FGSM achieve average white-box ASRs of 58.31\% and 91.00\%, respectively.
In contrast, MIG attains a white-box ASR of 93.48\%, while MIG-COW achieves 99.96\%, demonstrating its superior effectiveness. As expected, MIG-COW achieves the highest ASR by leveraging both consensus and model-specific complementary features through its COW mechanism. 

In the second setting, the ViT-P (a transformer-based architecture) is included as an additional white-box detector. However, in this setting, PGD and MI-FGSM exhibit limited generalization to the transformer architecture, achieving notably low ASRs of 25.74\% and 51.53\% on ViT-P, respectively. Meanwhile, MIG and MIG-COW maintain high performance, with ASRs of 90.96\% and 93.87\% on ViT-P. Notably, MIG-COW still achieves 99.93\% ASR on both CNN-based models, despite the inclusion of ViT-P. 
One potential reason for the baseline MIG's lower ASR on ViT-P compared to CNNs, despite its overall strength, is that ViT-P, due to its relatively poorer baseline performance on the AADD-2025 dataset (approximately 70\% accuracy compared to near-perfect accuracy for other models), may exhibit different integrated gradients from the perspective of output equivalence (c.f., implementation invariance). Our results suggest that MIG-COW effectively enhances the robustness of integrated gradients, even when dealing with a classifier that has a relatively poorer baseline performance.

\begin{figure*}[t]
  \centering
  \begin{subfigure}[t]{0.32\linewidth}
    \centering
    \includegraphics[width=\linewidth]{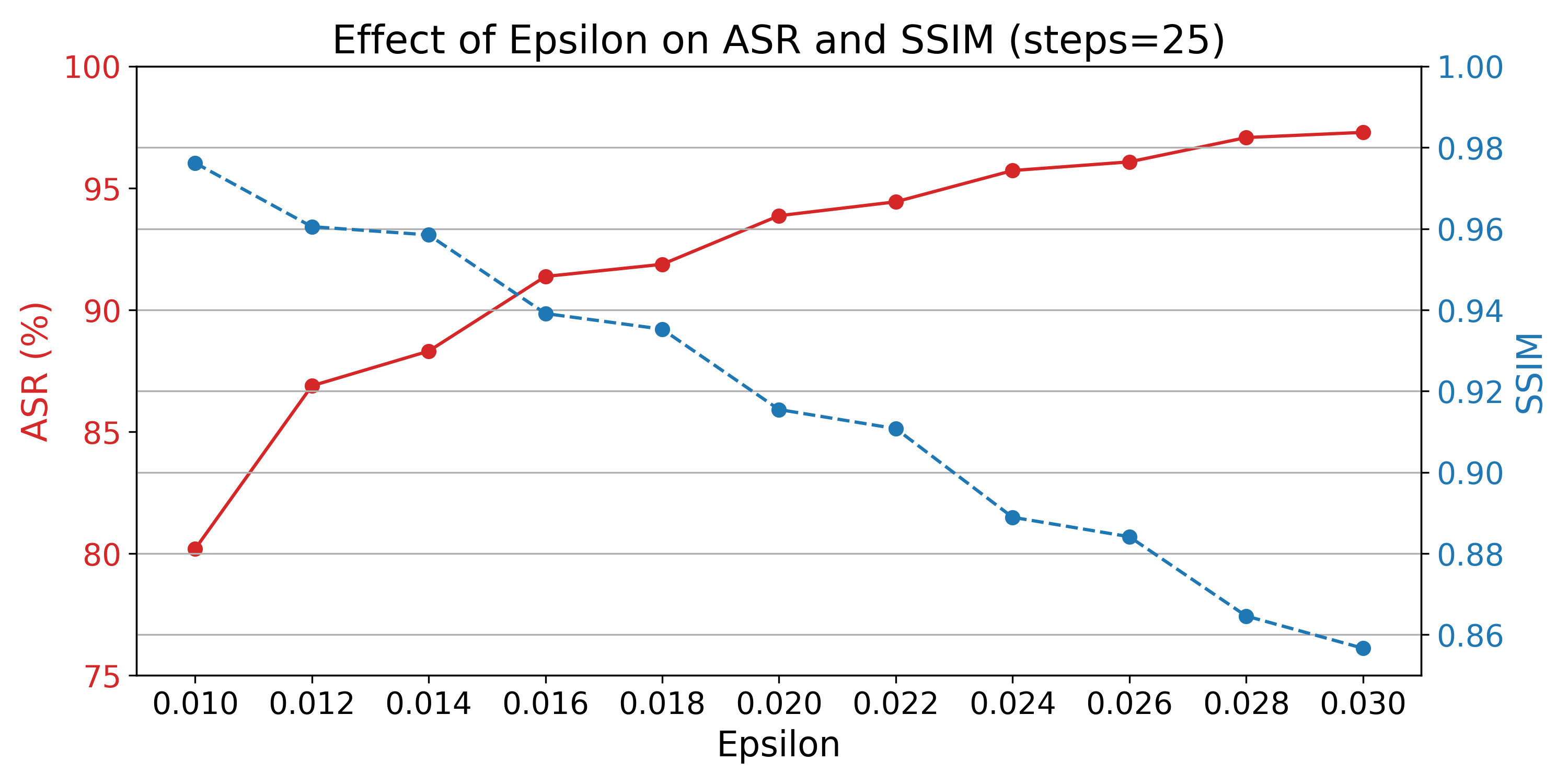}
    \caption{Tuning $\epsilon$}
    \label{fig:epsilon_tuning}
  \end{subfigure}
  \hfill
  \begin{subfigure}[t]{0.32\linewidth}
    \centering
    \includegraphics[width=\linewidth]{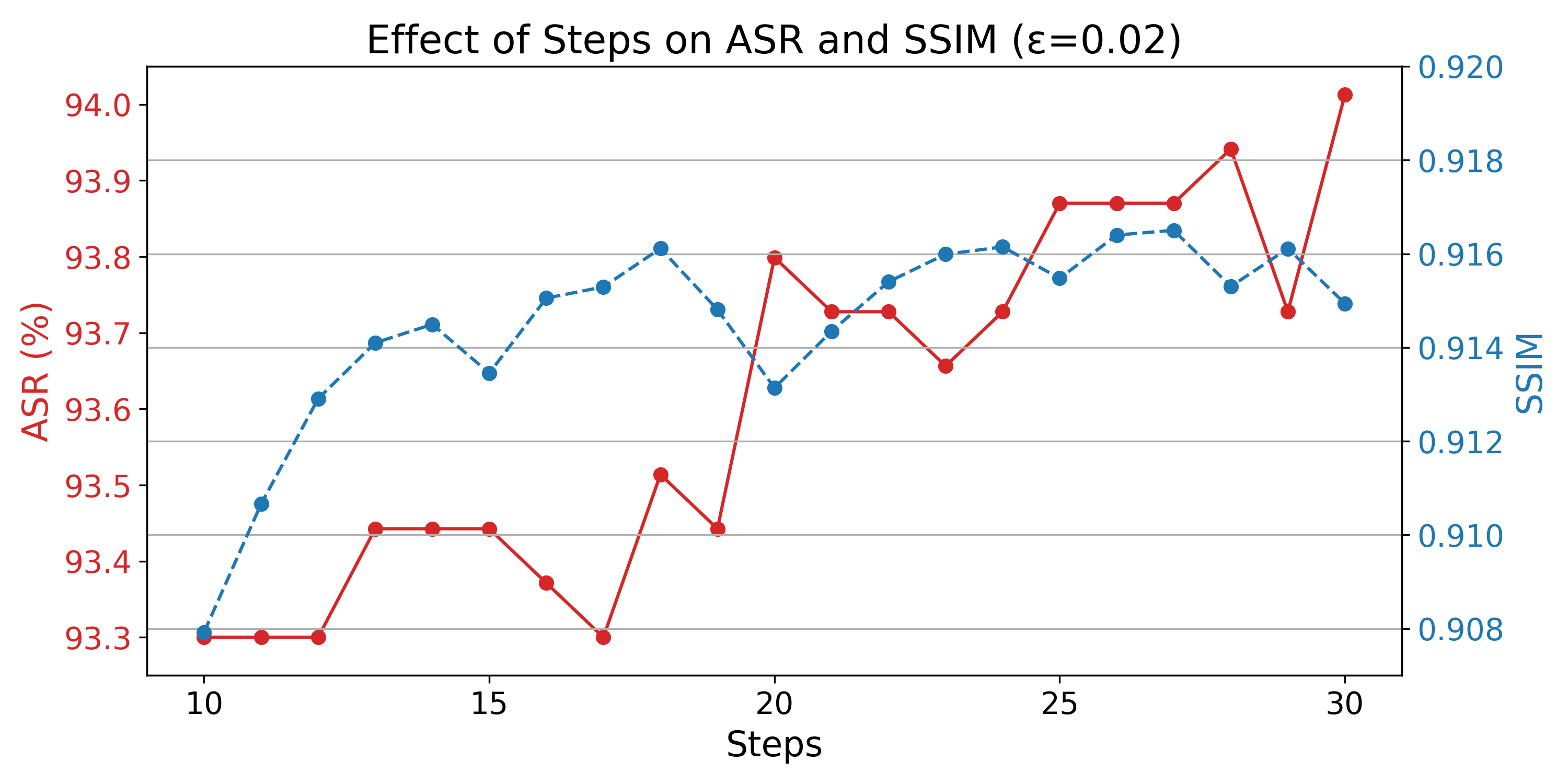}
    \caption{Tuning $T$}
    \label{fig:steps_tuning}
  \end{subfigure}
  \hfill
  \begin{subfigure}[t]{0.32\linewidth}
    \centering
    \includegraphics[width=\linewidth]{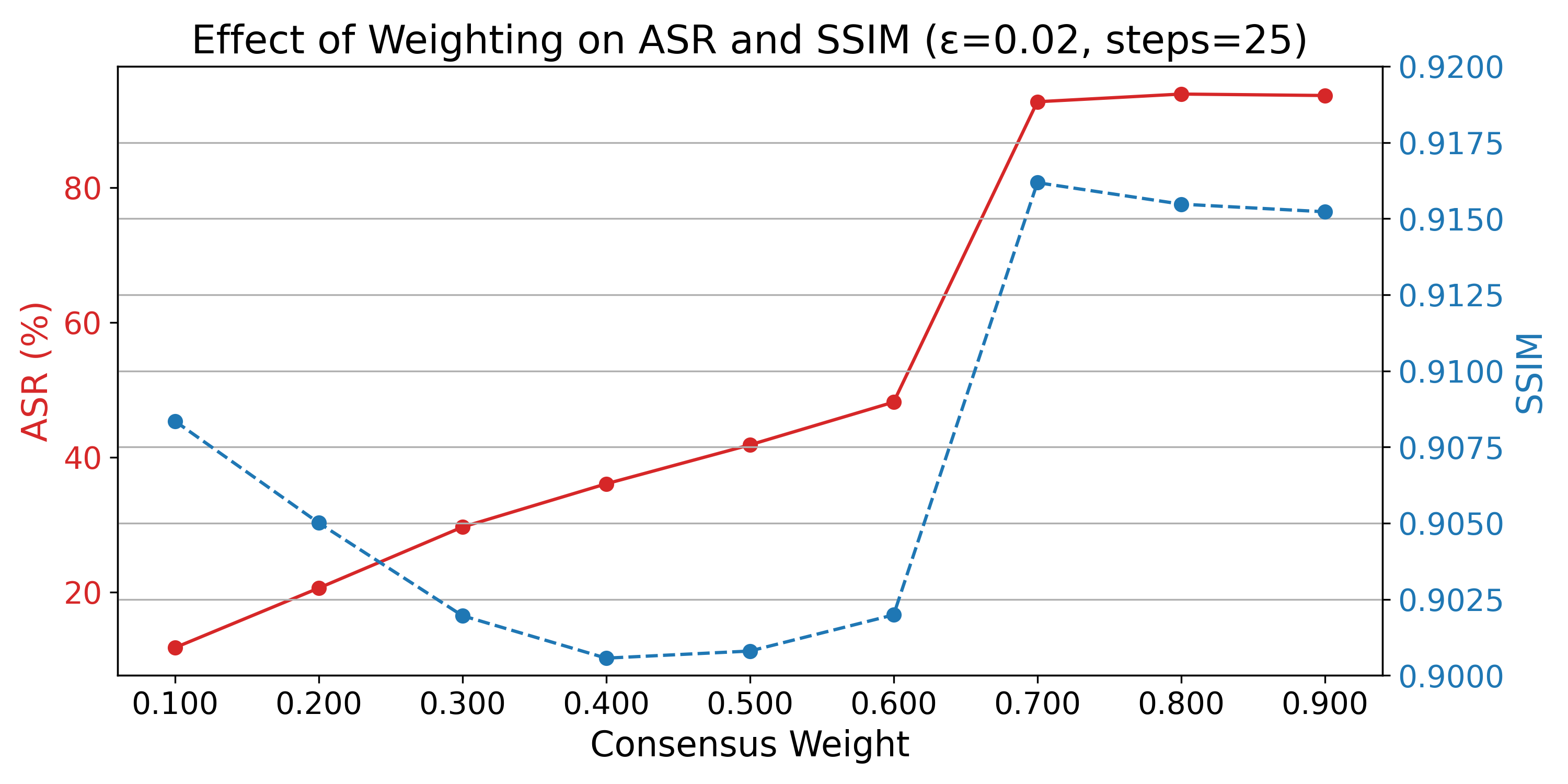}
    \caption{Tuning consensus weight}
    \label{fig:cw_tuning}
  \end{subfigure}
  \caption{Hyperparameter tuning results across perturbation budget $\epsilon$, steps $T$, and consensus weight $\mathbf{g}_\text{con}$ for ASR(\%) and SSIM, which are plotted as solid red and dashed blue lines, respectively, in each graph.}
\end{figure*}

\subsubsection{Attack on Black-box Detectors}\label{sec:black_att}
Our black-box evaluation strictly follows the AADD-2025 Challenge threat model, targeting official ViT-B/16 and DenseNet121-DCT models. The adversary has no access to model architecture, parameters, training data, or gradients, and is limited to a single adversarial query per input—receiving only a binary prediction result at test time, without confidence scores. Given these constraints, we employ a transfer-based strategy: adversarial examples are crafted using white-box models and evaluated on the black-box targets to measure transferability.

In this setting, MIG-COW achieves the highest average black-box ASR among all methods, while also delivering significantly better white-box ASR and SSIM. As a result, MIG-COW attains the best overall score by effectively balancing transferability, robustness, and imperceptibility.
However, similar to the white-box scenario, including ViT-P as a source model generally degrades black-box ASRs for IG-based methods. In this case, MI-FGSM achieves the highest black-box ASR, but its performance remains poor in the white-box setting.
Especially, for both MIG and MIG-COW, black-box ASR was reduced when ViT-P is included, suggesting that the baseline accuracy of source models plays a critical role in transfer-based attacks. Although ViT-P shares the same architectural structure (ViT-B/16) as one of the black-box targets, the results indicate that functional similarity—rather than structural alignment alone—is more important for achieving effective transferability.

\begin{table}[t]
    \centering
    \resizebox{\linewidth}{!}{
    \begin{tabular}{cc|ccc|c|c}
        \toprule
        \multicolumn{2}{c|}{Setting} & \multicolumn{3}{c|}{Detector} & \multirow{2}{*}{SSIM} & \multirow{2}{*}{Score} \\
        \cmidrule(lr){1-2} \cmidrule(lr){3-5}
        COW & CE loss & ResNet50 & DenseNet121 & ViT-P & \\
        \midrule
        \ding{51} &            & 92.37 & 93.08 & 94.51 & 0.9172 & 3,602 \\
                  & \ding{51}  & 100 & 99.93 & 89.02 & 0.9168 & 3,716 \\
        \ding{51} & \ding{51}  & 99.93 & 99.93 & 93.87 & 0.9161 & \textbf{3,772} \\
        \bottomrule
    \end{tabular}
    }
    \caption{White-box ablation study. Each detector score indicates performance under different CE loss and COW configurations. The Score is calculated based on the three models presented in the table.}
    \label{tab:whitebox_ablation}
\end{table}

\subsection{Component and Hyperparameter Analysis}
\subsubsection{Ablation Study}
We conduct an ablation study to evaluate the contribution of key components in the proposed MIG-COW framework: (i) replacing the logit-based cross-entropy loss (CE loss) with the original probability-based formulation used in MIG, and (ii) the inclusion of the orthogonal component from the consensus-orthogonal weighting (COW) strategy. These experiments aim to assess how each component impacts the generalizability and robustness of adversarial perturbations.

As shown in Table~\ref{tab:whitebox_ablation}, removing either the CE loss or the COW module consistently leads to evident performance drops in at least one detector. Specifically, excluding CE loss significantly reduces attack success on CNN-based models (ResNet50 and DenseNet121), while removing the COW module causes a moderate drop on ViT-P. This indicates that the CE loss provides more stable gradient signals for CNNs, while the COW module is critical for capturing model-specific features relevant to ViT-based detectors. MIG-COW—using both components—achieves consistently high performance across all models, with the highest overall score of 3,772. 

These results demonstrate that the CE loss and the COW module play complementary roles in balancing generalization and specificity. In particular, the orthogonal component of COW is vital for capturing model-specific decision boundaries not solely reflected in the consensus gradient, enabling simultaneous deception of detectors with diverse underlying architectures.

\subsubsection{Hyperparameters}
We conduct a sensitivity analysis to evaluate the effect of key hyperparameters on attack performance, focusing on: (1) the perturbation budget $\epsilon$, (2) the number of attack steps $T$, and (3) the consensus weight in our proposed gradient weighting method. 
For $\epsilon$, we vary its value from 0.01 to 0.03 (step size 0.002), keeping $T = 25$ fixed. As depicted in Figure~\ref{fig:epsilon_tuning}, increasing $\epsilon$ generally improves ASR but concurrently degrades SSIM, highlighting the inherent trade-off between attack strength and visual imperceptibility.
We also vary $T$ from 10 to 30 (step size 1), with $\epsilon = 0.02$ fixed to explore its impact. Figure~\ref{fig:steps_tuning} shows that increasing $T$ yields gradual improvements in both ASR and SSIM, with diminishing returns observed beyond $T = 25$.
To optimize the gradient weighting, we vary the consensus weight $\beta$ from 0.1 to 0.9 (step size 0.1), with the orthogonal weight set to $1 - \beta$, while keeping $\epsilon = 0.02$ and $T = 25$ fixed.
Figure~\ref{fig:cw_tuning} indicates that a consensus weight in the range of 0.7 to 0.8 maximizes ASR while preserving high perceptual similarity, suggesting an optimal balance between shared and model-specific gradient directions.
\section{Conclusion}
In this paper, we propose MIG-COW, a novel adversarial attack framework designed to enhance ensembling across diverse deepfake detection architectures. By leveraging momentum-based optimization alongside a consensus-orthogonal gradient decomposition, MIG-COW effectively captures both shared and model-specific vulnerabilities in heterogeneous ensembles of CNNs and ViT. Our experiments demonstrate MIG-COW's high attack success rates in white-box settings while preserving visual similarity with benign input. However, its reduced effectiveness against the official black-box ViT highlights a key challenge in real-world adversarial settings: bridging the gap between attacker-side surrogate models and unseen, deployment-time detectors with differing architectures, training objectives, and data distributions. We also show that incorporating a low-performing model into the white-box ensemble—despite increasing architectural diversity—can degrade overall attack performance in both white-box and black-box settings. These findings underscore the importance of carefully selecting surrogate models and motivate future research into adaptive, robust attack strategies. Addressing these challenges is essential for advancing adversarial machine learning and for more rigorous evaluations of the robustness of deployed deepfake detection systems.

\newpage

\begin{acks}
This work was supported by Institute of Information \& communications Technology Planning \& Evaluation (IITP) grant funded by the Korea government (MSIT) (No. RS-2020-II201336, Artificial Intelligence Graduate School Program (UNIST), 20\%) and ITRC (Information Technology Research Center) grant funded by the Korea government(Ministry of Science and ICT) (IITP-2025-RS-2024-00436936, 50\%). This work was also supported by the National Research Foundation of Korea (NRF) grant funded by the Korea government (MSIT) (RS-2025-00515481, 30\%). 
\end{acks}

\bibliographystyle{ACM-Reference-Format}
\balance
\bibliography{references.bib}
\end{document}